\documentclass[fleqn,twoside]{article}
\usepackage{espcrc2}
\usepackage{amsmath}

\usepackage{graphicx}

\newcommand{\ccbar}{\ensuremath{c\overline{c}}}
\newcommand{\dbar}{\ensuremath{\overline{D}}}
\newcommand{\epem}{\ensuremath{e^+ e^-}}
\newcommand{\ks}{\ensuremath{K_S^0}}

\newcommand{\dz}{\ensuremath{D^0}}
\newcommand{\dzbar}{\ensuremath{\overline{D}{}^0}}
\newcommand{\dstarb}{\ensuremath{\overline{D}{}^{\ast}}}
\newcommand{\dstarzb}{\ensuremath{\overline{D}{}^{\ast 0}}}
\newcommand{\pipi}{\ensuremath{\pi^+\pi^-}}
\newcommand{\ppbar}{\ensuremath{p\overline{p}}}
\newcommand{\qqbar}{\ensuremath{q\overline{q}}}

\newcommand{\ipb}{\ensuremath{\mathrm{pb}^{-1}}}
\newcommand{\iab}{\ensuremath{\mathrm{ab}^{-1}}}
\newcommand{\kev}{\ensuremath{\mathrm{keV}}}
\newcommand{\mev}{\ensuremath{\mathrm{MeV}}}

\newcommand{\pb}{\ensuremath{\mathrm{pb}}}

\newcommand{\br}{\ensuremath{\mathcal{B}}}

\newcommand{\mrecoil}{\ensuremath{M_{\text{recoil}}}}
\newcommand{\slj}[3]{\ensuremath{{}^{#1}{#2}_{#3}}}

\title{New charmonium-like states}

\author{B.D.~Yabsley\address[sydney]{High Energy Physics Group,	
					School of Physics, A28		\\
					University of Sydney. NSW 2006,	
					Australia}}

\begin{document}

\begin{abstract}
In recent years the $B$-factories and other machines have found
evidence for a large number of new states with hidden charm:
candidate $h_c(1P)$, $\eta_c(2S)$, and $\chi_{c2}(2P)$ states; the
well-established $X(3872)$; enhancements called $X(3940)$,
$Y(3940)$, and $Y(4260)$; and a new structure at 4350 MeV. Various
conventional-charmonium and more exotic interpretations of these
data have been proposed. In this talk we review the current state
of the experimental evidence and the prospects for clarifying the
spectrum.
\vspace{1pc}
\end{abstract}

\maketitle

\section{INTRODUCTION}

For many years it was possible to consider the charmonia 
as a ``well-understood'' system: the spectrum of low-lying states was
well-established, with one or two gaps, as were the major transitions.
However recently there has been a long list of discoveries,
or claims of evidence:
the $h_c$, excitation(s) of the $\eta_c$,
and an alphanumeric soup of ``X'' and ``Y'' mesons
(and briefly a ``Z''),
some understandable as \ccbar\ states,
some where a conventional assignment is elusive.
This revival in spectroscopy is due in part to the large datasets
accumulated at the $B$-factories,
and (one assumes) to the lack of strong new-physics-in-$CP$-violation signals
there, making analysis of hidden-charm states respectable.  
But much of the interest has been due to the data unsettling our understanding
of this sector.
At least two of the new charmonium-\emph{like} states
are candidates for non-\qqbar\ mesons,
and there are other features in the data that are not yet understood.
In the following we will briefly survey the recent results,
with some emphasis on these suggestive cases.

\section{THE $X(3872)$}

The most extensively studied of the new states is the so-called
$X(3872)$. Its existence is well-established,
and some properties have become increasingly clear;
its mass suggests a bound molecule-like state of $D^0$ and \dstarzb,
but this interpretation is by no means uncontested.
%A summary is beyond the scope of this talk. 

\subsection{Discovery \& charmonium exclusions}

The $X(3872)$ was found by Belle~\cite{x3872-belle-discovery},
with a clear signal in decays $B^+ \to K^+ X[\to \pipi J/\psi]$. 
Confirmations followed in short
order~\cite{x3872-cdf-confirmation,x3872-d0-confirmation,x3872-babar-confirmation}.
The observations agree on a peak with natural width below experimental
resolution,
and a dipion mass spectrum concentrated at high values, 
favouring $X \to \rho \psi$ and hence $C$-parity $\xi_C = +1$.
Decays to $D\dbar$ are not seen~\cite{x3872-belle-chistov}.
Together with the narrowness of the state,
this suggests such decays are forbidden,
disfavouring $J^{PC} = 0^{++},\,1^{--},\,2^{++}$ \emph{etc}.
Thus a conventional-charmonium $X$ would likely have unnatural $J^P$ 
(the remaining assignments $0^{+-},\,1^{-+},\,\ldots$ are exotic).
The corresponding $J\leq 2$ candidates were surveyed
for MESON 2004~\cite{x3872-olsen-exclusions};  
$\psi_3(1\slj{3}{D}{3})$ was also included,
as decay to $D\dbar$ would be suppressed by angular momentum in this case.
There are arguments against each assignment: \emph{e.g.}\ Belle finds 
$\Gamma(X\to\gamma\chi_{c1})/\Gamma(X\to\pi\pi\psi) < 0.89$
at 90\% confidence, while this ratio is expected to be $>1.6$
for the $\psi_2(1\slj{3}{D}{2})$.
%[potential $/$ $\psi''$ Wigner-Eckart]
This, and some other tests, 
would bear repetition with the larger event samples currently available.

\subsection{Determination of quantum numbers}

\begin{figure}
  \includegraphics[width=7cm]{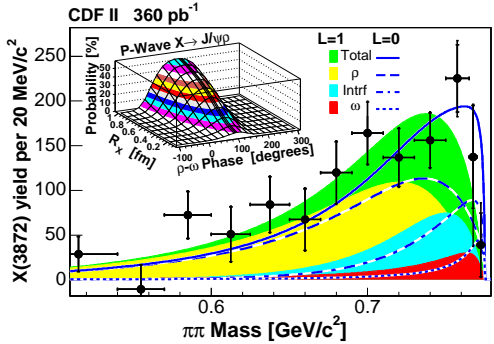}
  \caption{CDF~\cite{x3872-cdf-dipion}: detail of the dipion mass spectrum
	and fits to $X(3872) \to \rho\psi$ models with $L=0,\,1$.
	%including $\rho$, $\omega$, and $\rho$-$\omega$ interference terms.
	Inset: $L=1$ fit probabilities \emph{vs}
	$\rho$-$\omega$ interference phase and radius of interaction $R_X$.}
  \label{fig-x3872-cdf-dipion}
\vspace*{-0.8ex}
\end{figure}

The even $C$-parity of the $X$ was fixed by Belle's observation
of the radiative transition to $J/\psi$ ~\cite{x3872-belle-radiative},
with $\Gamma_{\gamma\psi}/\Gamma_{\pipi\psi} = 0.14 \pm 0.05$,
confirmed by BaBar~\cite{x3872-babar-radiative}
with a larger rate.
Belle also sees
$X\to \omega J/\psi$, nominally sub-threshold,
\emph{via} the low-mass tail of the $\omega$, 
finding $\br_{\pipi\pi^0\psi}/\br_{\pipi\psi}=1.0\pm 0.4\pm 0.3$~\cite{x3872-belle-radiative};
this likewise requires $\xi_C = +1$.

Belle's angular analysis~\cite{x3872-belle-angular} 
favours $J^{PC} = 1^{++}$ over most other quantum numbers;
$2^{++}$ and $2^{-+}$ are not excluded.
A beautiful analysis of the dipion mass spectrum by CDF~\cite{x3872-cdf-dipion}
rules out \slj{1}{P}{1} ($h_c^\prime$) and \slj{3}{D}{J} ($\psi_J$)
charmonium assignments; \slj{3}{S}{1}\ states are allowed,
but if we identify $\psi(4040) \equiv \psi(3S)$, none are available. 
Only $X\to \rho\psi$ remains, 
with fits to both $L=0$ and 1 allowed by the data (Fig.~\ref{fig-x3872-cdf-dipion}).
This study tends to supersede a Belle fit that found $L=1$
(and hence, odd-parity) solutions were disfavoured~\cite{x3872-belle-angular}.
If CDF mimic Belle's fit by excluding Blatt-Weisskopf form factors
and $\rho$-$\omega$ interference, they find similar results.
Since the conference, CDF have submitted their own angular analysis
for publication~\cite{x3872-cdf-angular}: 
only $1^{++}$ and $2^{-+}$ are permitted.

Belle sees a peak at $\dz\dzbar\pi^0$ threshold
in $B\to \dz\dzbar\pi^0 K$~\cite{x3872-belle-ddpi}, 
disfavouring $2^{++}$ if interpreted as an $X$ signal.
The implied decay rate is an order of magnitude larger
than that to $\pipi\psi$. The energy release is 
$Q = 11.2\,\mev$,
corresponding to $M=(3875.9 \pm 0.7 ^{+0.3}_{-1.6} \pm 0.4)\,\mev$
(using CLEO's \dz-mass~\cite{dmass-cleo}).
This is well above the $X$ mass~\cite{pdg2006}.
Confirmation and further study are needed.

\subsection{Searches for partner states}

%The $X(3872)$ was originally observed in $B^+\to K^+ X$.
If the $X$ is a \dz\dstarzb\ molecule, and produced \emph{via} 
$B\to K \dz\dstarzb$, the rate in $B^0$ decays ($\br^0$) should be suppressed
with respect to the rate in $B^+$ decays ($\br^+$) by a ratio 
$R \equiv \br^0/\br^+ <0.1$~\cite{x3872-braaten-kusunoki}.
In another model the $X$ is one of a family of 
diquark-antidiquark states~\cite{x3872-maiani},
$X_u = [cu][\bar{c}\bar{u}]$ or $X_d = [cd][\bar{c}\bar{d}]$,
or a mixture; $M(X_d) - M(X_u) \approx (7\pm 2)\,\mev$.
Here $X_u$ or $X_d$ would dominate the  $B^+\to K^+ X$ peak,
with its partner contributing at the same rate ($R=1$) in $B^0$ decay.
BaBar~\cite{x3872-babar-xu-xd} finds
a $2.5\sigma$ $B^0 \to \ks\,X$ signal,
with $R = 0.50\pm 0.30\pm 0.05$ ($R\in[0.13,1.10]$ at 90\% C.L.),
and $\Delta M =  M^+ - M^0 = (2.7\pm 1.3 \pm 0.2)\,\mev$.
The mass difference is indecisive (it vanishes for \dz\dstarzb),
but $R$ weakly favours $qq\bar{q}\bar{q}$ over the molecular model.
Further $\dz\dzbar\pi^0$ and $\dz\dzbar\gamma$ study (as above)
might be helpful here~\cite{x3872-braaten-kusunoki}.

BaBar has also searched for a charged partner
appearing as a $\pi^-\pi^0\psi$ mass peak in 
$B\to K\pi^-\pi^0\psi$~\cite{x3872-babar-charged}.
No signal is seen and the resulting upper limits disfavour
an isovector $X(3872)$.

\subsection{Interpreting the $X(3872)$: a summary}

The discovery mode $X\to \pipi\psi$ is prominent,
with branching fraction $> 4.2\%$ (at 90\% confidence)
based on BaBar's study of the inclusive kaon momentum spectrum
in $B^\pm$ decay~\cite{x3872-babar-inclusive}.
The mass $(3871.2 \pm 0.5)\,\mev$
is now \emph{below} the $\dz\dstarzb$ threshold of $(3871.81\pm 0.36)\,\mev$
set by CLEO's post-conference measurement~\cite{dmass-cleo}.
%	the binding energy of a \dz\dstarzb\ molecule
%	at that mass would be $(0.6\pm 0.6)\,\mev$
The width is still unresolved, with no advance on the 
limit $\Gamma_X < 2.3\,\mev$ (90\% C.L.) from Ref.~\cite{x3872-belle-discovery}.

The isospin violation seen in the $X(3872)$ decay to both
$\rho\psi$ and $\omega\psi$ is natural for a \dz\dstarzb\ molecule,
as are the mass and the lack of charged partners;
no data conflict seriously with this hypothesis.
CDF~\cite{x3872-bauer} finds $X$ production in \ppbar\ 
to be $\psi(2S)$-like, which tends to imply a compact \emph{component},
say \ccbar; this might account for the tantalising
$B^0$-\emph{vs}-$B^+$ production results~\cite{x3872-babar-xu-xd}.
While calls for ``more data'' from the $B$-factories (at $1\,\iab$)
are tedious, more precise $R$ and $\Delta M$ are needed.

Only $J^{PC} = 1^{++}$ or $2^{-+}$ are plausible,
but the corresponding charmonium assignments
$\chi_{c1}^\prime$~\cite{review-elq} and 
$\eta_{c2}$ ($1\slj{1}{D}{2}$)~\cite{x3872-olsen-exclusions}
% $\Gamma(\eta_{c2} \to \pipi \eta_c)$ (I-cons.) sh$^d$ be $\gg \Gamma(\eta_{c2}\to\pipi\psi)$;
%  \& $\eta_{c2} \stackrel{E1}{\not\!\!\longrightarrow} \gamma\psi$
are not;
$\gamma\psi'$ and $\pipi\eta_c$ results would be decisive for these cases.
A more serious loose end is $\pi^0\pi^0\psi$,
although the $X\to\rho[\to\pipi]\psi$ interpretation 
(excluding all charmonia) is already favoured;
better understanding of $X\to\omega\psi$
(the underlying amplitude must be large),
and $D\dbar\{\gamma,\pi^0\}$ studies,
are also desirable.

\section{$h_c(1\slj{1}{P}{1})$: FINALLY OBSERVED}

\begin{figure}
  \includegraphics[width=7cm]{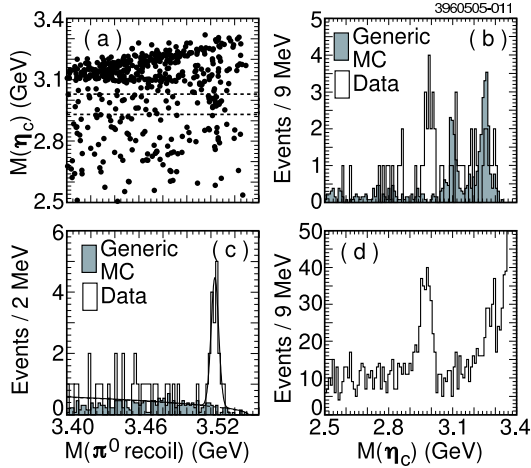}
  \caption{CLEO~\cite{hc-cleo} 
	$\psi(2S) \to \pi^0 h_c$, $h_c\to \gamma \eta_c$ analysis,
	with $\eta_c$ reconstruction:
	(a) $M(\eta_c)$ \emph{vs} $\pi^0$ recoil mass,
	(b) $M(\eta_c)$ and (c) $\mrecoil(\pi^0)$ projections
	under cuts on the other variable, and
	(d) signal for the normalisation mode $\psi(2S) \to \gamma \eta_c$.}
  \label{fig-hc-cleo-exclusive}
\vspace*{-0.3cm}
\end{figure}

CLEO~\cite{hc-cleo} has observed $h_c\to\gamma\eta_c$,
confirming weaker evidence from E835~\cite{hc-e835} in the same mode.
Using $3.09\times 10^6$ $\epem\to\psi(2S)$ events,
CLEO reconstruct the isospin-violating decay $\psi(2S)\to\pi^0 h_c$. %;
%a veto on $\pi^0$ from $J/\psi\to\pi^0 X$ and $\psi(2S)\to\pi^0\pi^0J/\psi$
%is applied.
One analysis proceeds by $\eta_c$ reconstruction,
and a kinematically constrained fit to the event; the  
results are shown in Fig.~\ref{fig-hc-cleo-exclusive}.
%(The $h_c$ mass is recovered from the $\psi(2S)$ and $\pi^0$ four-momenta
%to improve the resolution.)
An inclusive analysis (without $\eta_c$ reconstruction) is also performed,
with compatible results; the helicity angle distribution of the photon
is consistent with $(1+\cos^2 \theta)$ [$\chi^2/n_{dof}=1.7/2$],
as expected for $h_c\to\gamma\eta_c$.

The product branching fraction
$\br_{\psi(2S)}\times\br_{h_c} = (4.0\pm 0.8 \pm 0.7)\times 10^{-4}$
bisects a range of theoretical estimates covering
two orders of magnitude;
the mass $M(h_c) = (3524.4\pm 0.6\pm 0.4)\,\mev$
yields a hyperfine splitting  
$\langle M(\slj{3}{P}{J}) \rangle - M(\slj{1}{P}{1}) = (+1.0\pm 0.6 \pm 0.4)\,\mev$, consistent with expectations
(the spin-spin interaction vanishes at lowest order).
E835 quote $M(h_c) = (3525.8\pm 0.2 \pm 0.2)\,\mev$, somewhat higher
and more precise, and a limit $\Gamma(h_c) < 1\,\mev$ on the width.

\section{$\eta_c(2\slj{1}{S}{0})$ AND THE $X(3940)$ ($3\slj{1}{S}{0}$?)}

\begin{figure}
  \includegraphics[width=6cm]{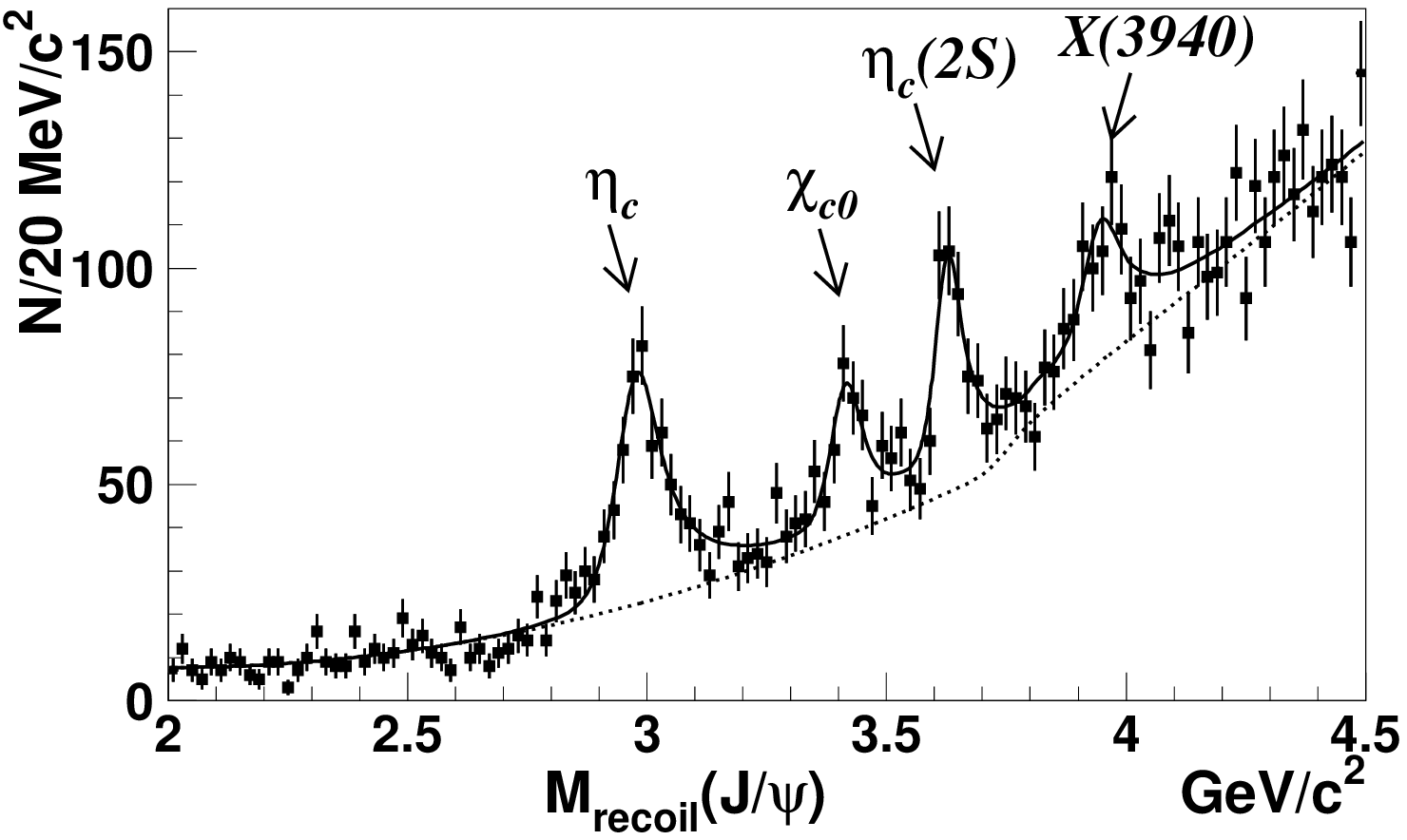}
  \\[0.5cm]
  \includegraphics[width=6cm]{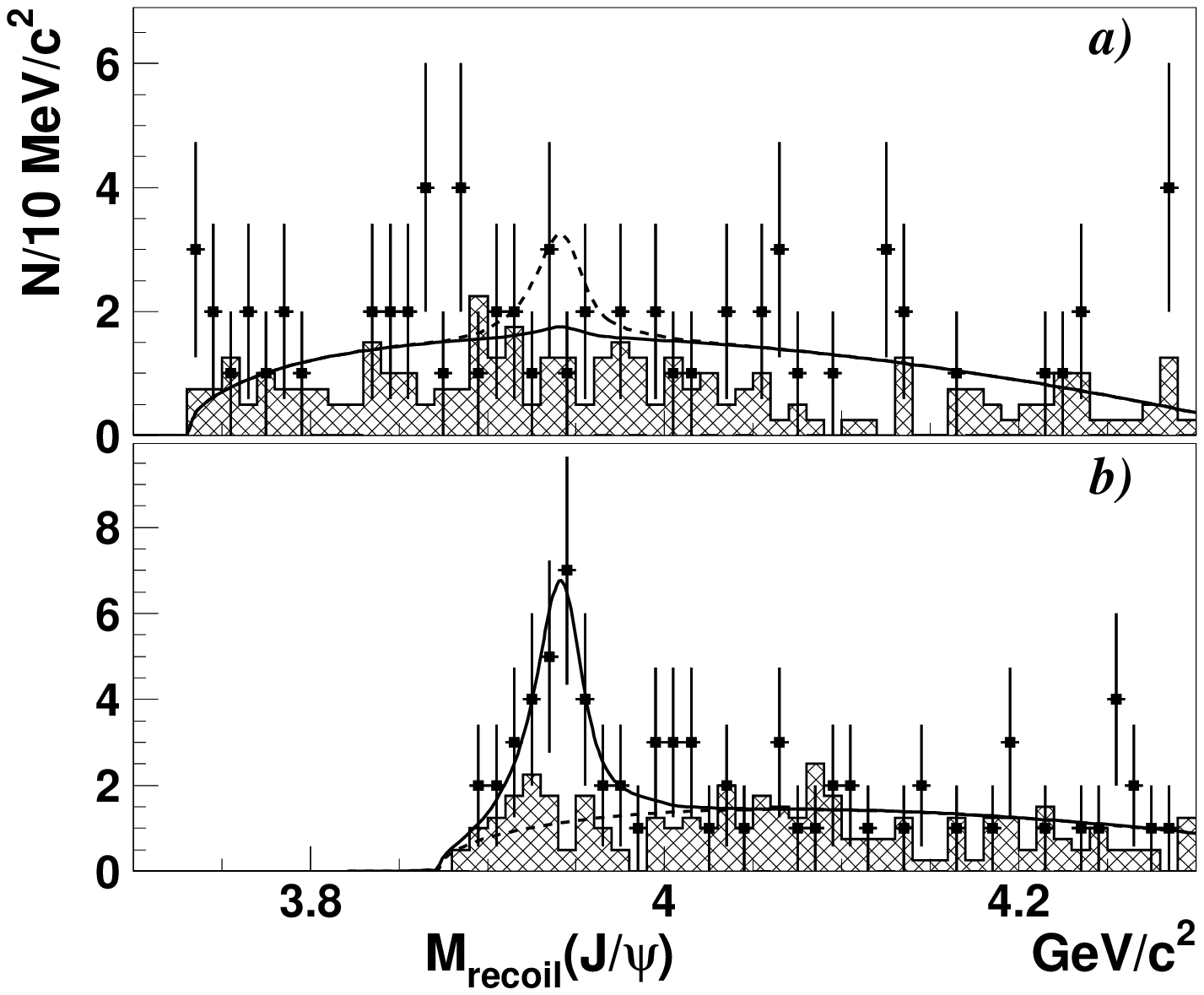}
  \caption{Belle~\cite{x3940-belle}:
	$\mrecoil(J/\psi)$ distributions
	(upper) for $\epem \to J/\psi\,X$, 
	(lower) for events tagged and constrained as
		(a) $\epem\to J/\psi\,D\dbar$ and
		(b) $\epem\to J/\psi\,D\dstarb$.}
  \label{fig-belle-x3940}
\vspace*{-0.3cm}
\end{figure}

Belle's unexpected double charmonium production results~\cite{psi-etac-belle}
have now been confirmed by BaBar~\cite{psi-etac-babar}: both groups
see large $\epem\to J/\psi\,\eta_c$ and $J/\psi\,\eta_c^\prime$ rates.
The $\eta_c^\prime$ is now well-established, 
with observations in $B\to K\eta_c^\prime$~\cite{etac2s-belle} and
$\gamma\gamma\to\eta_c^\prime$~\cite{etac2s-cleo,etac2s-babar}.
The average~\cite{pdg2006} width $\Gamma = (14\pm 7)\,\mev$,
and mass $M=(3638\pm 4)\,\mev$; %(scale factor $1.8$!);
the hyperfine splitting $M_{\psi(2S)}-M_{\eta_c(2S)} \sim 48\,\mev$
is somewhat smaller than theoretical estimates~\cite{review-elq}.

Belle~\cite{x3940-belle} also finds a significant peak
above open-charm threshold, the ``$X(3940)$'' 
(see Fig.~\ref{fig-belle-x3940}).
$D\dbar{}^{(*)}$ decays are studied by reconstructing a $D$-meson,
and constraining $\mrecoil(J/\psi D)$ to the $D^{(*)}$ mass;
a clear $X\to D\dstarb$ peak is seen in $\mrecoil(J/\psi)$,
but none for $D\dbar$. 
Belle finds $M_X = (3943\pm 6 \pm 6)\,\mev$ and
$\Gamma_X = (15.4\pm 10.1)\,\mev$
($<52\,\mev$ at 90\% C.L.).
$D\dstarb$ dominates among final states with more than two tracks:
$\br_{>2}(X\to D\dstarb) > 45\%$ at 90\% C.L.;
$\br(X\to D\dbar) < 41\%$.
%Based its production, width, and decay modes, 
Belle speculates $X(3940) \equiv \eta_c(3S)$, implying a 
large hyperfine splitting~\cite{review-elq}.
Angular analysis~\cite{review-barnes}, or
$\gamma\gamma\to X\to D\dstarb$~\cite{review-godfrey},
could confirm this assignment.
A $\chi_{c0}(2P)$ contribution to the $J/\psi\,X$ peak is not excluded,
but $\chi_{c0}(2P) \not\to D\dstarb$;
$X = \chi_{c1}(2P)$ seems unlikely,
as no $J/\psi\,\chi_{c1}(1P)$ signal is seen.

\section{THE ``$Y(3940)$'' AT $\omega\, J/\psi$ THRESHOLD}

Belle~\cite{y3940-belle} also see an $\omega J/\psi$ peak at similar mass,
\emph{i.e.}\ just above threshold, in $B \to K \omega J/\psi$.
Fitting with a single $S$-wave Breit-Wigner,
they find a significant peak above phase space,
with $M = (3943 \pm 11 \pm 13)\,\mev$ and $\Gamma = (87 \pm 22 \pm 26)\,\mev$,
denoted $Y(3940)$.
The product branching fraction
$\br(B\to K Y)\times\br(Y\to\omega J/\psi)=(7.1\pm 1.3 \pm 3.1)\times 10^{-5}$
implies large $\br(Y\to\omega J/\psi)$ if $Y$ is a charmonium,
but such a state would presumably decay to $D\dbar{}^{(*)}$,
with only small $\omega\psi$ branching.  
For comparison, from Ref.~\cite{x3872-belle-chistov},
$\br(B^+\to K^+ Y)\times\br(Y\to D\dbar) < O(5\times 10^{-5})$.
The $X(3940)$ decays to $D\dstarb$,
but $\br(X(3940)\to \omega J/\psi) < 26\%$ at 90\% C.L.~\cite{x3940-belle}.
If the $Y(3940)$ were the $\chi_{c1}(2P)$,
the large $\omega\psi$ branching might be explained by rescattering
from $D\dstarb$~\cite{review-godfrey,review-barnes};
there is no evidence yet that the $X(3940)$ is the same state.
%  \item $\br( B^+ \to K^+ Y)\times \br(Y\to\gamma J/\psi) < 1.4\times 10^{-5}$
%        \cite{x3872-babar-radiative}
The observed $Y(3940)$ decay would be consistent with expectations
for a $\ccbar g$ hybrid~\cite{y3940-belle}, but the mass would not.

\section{THE $Z(3930)$ IN $\gamma\gamma \to D\dbar$: $\chi_{c2}(2\slj{3}{P}{2})$?}

Belle's third structure in this mass region, the $Z(3930)$~\cite{z3930-belle},
is easier to interpret: it has already been acclaimed as the $\chi_{c2}(2P)$
by the PDG~\cite{pdg2006} without confirmation.
The $Z$ is a significant peak in $\gamma\gamma \to D\dbar$,
fitted by a relativistic Breit-Wigner with
$M = (3929 \pm 5 \pm 2)\, \mev$ and $\Gamma = (29 \pm 10 \pm 2)\,\mev$.
%The signal in $D^+ D^-$ is weak, but the resulting
The ratio
$\br(Z\to D^+ D^-) / \br(Z\to \dz\dzbar) = 0.74 \pm 0.43 \pm 0.16$ is consistent
with the value $0.89$ expected from isospin and the $D^+$-\dz\ mass difference.
Fitting the angular distribution of the $D$-mesons
relative to the \epem\ beam axis,
spin and helicity $(J,M) = (2,\, \pm 2)$ are strongly favoured over $J=0$, with
$\chi^2 = 1.9$ \emph{vs} $23.4$, for $n_{dof}= 9$.
Assuming spin-2,
$\Gamma_{\gamma\gamma} \times \br_{D\dbar} = (0.18 \pm 0.05 \pm 0.03)\,\kev$.
These results are consistent with expectations for the $\chi_{c2}(2P)$,
although the mass and $\Gamma_{\gamma\gamma} \times \br_{D\dbar}$
are on the low side~\cite{review-elq,review-godfrey,review-barnes};
a measurement of $\gamma\gamma\to Z\to D\dstarb$ would be useful confirmation.
%\emph{cf.}\ $\br(B^+\to K^+ Z) \times \br(Z\to \gamma J/\psi) < 2.5\times 10^{-6}$
%\cite{x3872-babar-radiative}

\section{$Y(4260)$: A HYBRID CANDIDATE}

\begin{figure}
  \includegraphics[width=7cm]{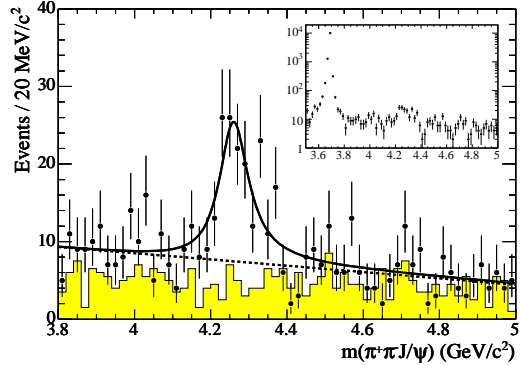}
  \caption{BaBar~\cite{y4260-babar-discovery}:
	$\pipi J/\psi$ invariant mass spectrum in initial state radiation
	events, with $J/\psi$ mass sidebands shaded, and the single-resonance
	($Y(4260)$) fit shown. A wider mass range is shown inset.}
  \label{fig-y4260-babar}
\vspace*{-0.3cm}
\end{figure}

BaBar's observation~\cite{y4260-babar-discovery} of the $Y(4260)$ 
generated great interest, because %(as with the $X(3872)$)
its mass and properties suggest an exotic structure:
a hybrid meson with valence partons $\ccbar g$.
In initial state radiation (ISR) events $\epem \to \gamma_{ISR} \pipi J/\psi$,
%BaBar see a prominent structure around 4260 MeV
%in the $\pipi J/\psi$ invariant mass spectrum (Fig.~\ref{fig-y4260-babar}).
BaBar see a prominent structure in the $\pipi J/\psi$ invariant mass spectrum
around 4260 MeV (Fig.~\ref{fig-y4260-babar}).
A state appearing in ISR must have $J^{PC} = 1^{--}$,
but a new vector charmonium at this mass is implausible:
candidates for the expected states are already known,
and decays should be dominated by $D^{(*)} \dbar{}^{(*)}$;
%with all three channels open;
the branching fraction to $\pipi\psi$ should be small. 
For a hybrid, however, decays
to a pair of $S$-wave mesons $D^{(*)}$ are highly suppressed~\cite{hybrid-page},
while decay to $\pipi\psi$ is plausible, as are a mass $\sim 4260\,\mev$ and
$J^{PC} = 1^{--}$~\cite{hybrid-close-page}.

BaBar fits using a relativistic Breit-Wigner, finding 
$M = (4259 \pm 8 ^{+2}_{-6})\,\mev$ and 
$\Gamma = (88 \pm 23 ^{+6}_{-4})\,\mev$,
and $125 \pm 23$ events;
they cannot rule out contributions by more than one state.
%The same distribution for $\pipi\ell^+\ell^-$ combinations with 
%$\ell^+\ell^-$ from the $J/\psi$ mass sidebands
%(also shown in Fig.~\ref{fig-y4260-babar}) is featureless;
Recoil mass and other distributions are consistent with ISR production.
$\Gamma_{\epem}\times\br_{\pipi\psi}=(5.5\pm 1.0 ^{+0.8}_{-0.7})\,\mathrm{eV}$,
implying a dielectron width too narrow for \ccbar,
if the $\pipi\psi$ branching is large. 

\begin{figure}
  \includegraphics[width=6cm]{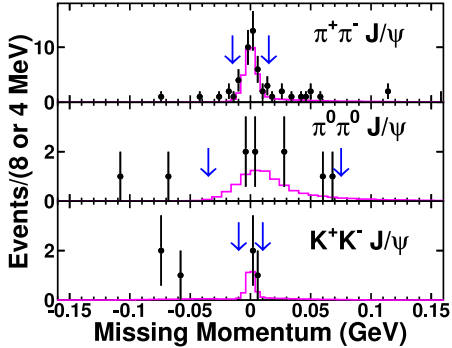}
  \caption{CLEO~\cite{y4260-cleo-isr}: missing momentum distributions
	in $\epem\to Y(4260)$ data.
	Curves show Monte Carlo expectation, scaled to the signal.}
  \label{fig-y4260-cleo-modes}
\end{figure}

CLEO~\cite{y4260-cleo-escan} has confirmed the $\pipi\psi$ signal 
using $13.2\,\ipb$ of \epem\ data at 4260 MeV;
the sample is very clean (Fig.~\ref{fig-y4260-cleo-modes}).
%with a 37 event signal
%on a $\sim 2.4$ event background (Fig.~\ref{fig-y4260-cleo-modes}).
Evidence is also seen for two new modes, with cross-sections of
$(23^{+12}_{-8} \pm 1)\,\pb$ for $\pi^0\pi^0\psi$ and
$(9^{+9}_{-5}   \pm 1)\,\pb$ for $K^+K^-\psi$, 
\emph{cf.}\ 
$(58^{+12}_{-10} \pm 4)\,\pb$ for $\pipi\psi$,
consistent with BaBar.
They note that these modes disfavour some other (non-hybrid)
exotic proposals~\cite{y4260-chi-rho,y4260-baryonium}; 
limits on decays of other vector charmonia to $\pipi\psi$ 
confirm that a conventional assignment is disfavoured
(\emph{cf.}\ Ref.~\cite{y4260-psi4S}). 
$O(10\,\text{\textendash}\,100\,\pb)$ limits are set on cross-sections for 
$\epem \to \{\pi^0,\,\eta,\,\eta',\,\pi^+\pi^-\pi^0,\,\eta\eta\}\, J/\psi$,
$\{\pi^+\pi^-,\,\eta\}\,\psi(2S)$,
$\omega \chi_{c0}$,
$\gamma \chi_{c1,c2}$,
$\pi^+\pi^-\pi^0 \chi_{c1,c2}$ and
$\pi^+\pi^- \phi$.
BaBar has also presented limits
on such modes~\cite{y4260-babar-confmodes}.

Belle~\cite{y4260-belle} confirmed the ISR signal
at summer conferences, finding
a $165 \pm 24^{+7}_{-23}$ event Breit-Wigner peak with
$M = (4295 \pm 10 ^{+10}_{-3})\,\mev$ and
$\Gamma = (133 \pm 26^{+13}_{-6})\,\mev$;
%both larger than BaBar's;
the recoil-mass and angular distributions are ISR-like.
(Since Beauty2006, CLEO~\cite{y4260-cleo-isr} has also published
supporting results in ISR. 
The peak mass $(4284^{+17}_{-16} \pm 4)\,\mev$
lies between the conflicting $B$-factory values.)

A lower limit $\br(Y\to\pipi\psi) > 0.6\%$
has been set~\cite{y4260-bes} based on the BES hadronic cross-section.
Studies of exclusive $D\dbar$~\cite{y4260-babar-isr-ddbar}
and $D^{(*)+}D^{*-}$~\cite{y4260-belle-isr-ddbar} spectra
in ISR at the $B$-factories likewise find no $Y$ signal;
this region is a local minimum in a complicated distribution.
A search is now needed for $Y\to D\dbar_1(2420)$,
which (for a hybrid) should dominate if allowed~\cite{hybrid-close-page};
with thresholds at 4287 and 4296 MeV, 
this question is tied to that of the mass (and shape) of the peak.
Related hybrid states, with other $J^{PC}$, might also be sought.

%\section{A structure at $4350\,\mev$?} 

BaBar showed~\cite{y4350-babar} a peak in ISR $\pipi \psi(2S)$
data at the summer,
said to be inconsistent with the $Y(4260)$, $\psi(4415)$,
or phase space; %assuming a single resonance 
they fit $M = (4354\pm 16)\,\mev$ and $\Gamma = (106\pm 19)\,\mev$. 
The influence on $Y(4260)$ results of coupling to channels near threshold
has already been discussed
in models with (\emph{e.g.}~\cite{hybrid-close-page})
and without a new state (\emph{e.g.}~\cite{y4260-coupled-channel}).
In the next generation of studies, more sophisticated fits in a
range of modes will presumably clarify
whether we are dealing with 0, 1, 2, or more new states.

\section{SUMMARY}

Becoming increasingly speculative as we progress:
the $h_c$ has been seen,
and the $\eta_c(2S)$ established;
the $Z(3930)$ is probably the $\chi_{c2}(2P)$,
and the $X(3940)$ may be the $\eta_c(3S)$.
%	although $2\slj{3}{P}{J}$ states might be confusing the picture.
Whatever the $X(3872)$ is, it looks like it has lots of $D\dstarb$
	in its wavefunction.
It's not clear what the ``$Y(3940)$'' is:
        confirmation of the structure would be helpful.
The $Y(4260)$ so far looks like a $\ccbar g$ hybrid.
More data are needed for the last three items,
and seem likely to be forthcoming.
Finally, something may be going on at $4350\,\mev$, but if so,
it is unlikely to be the last new development in the hidden-charm sector.

\end{document}